\documentclass[11pt]{article}

\usepackage{amsmath}
\usepackage{graphicx}
\usepackage{amssymb}
\usepackage{amsfonts}

\def\beq{\begin{eqnarray}}
\def\eeq{\end{eqnarray}}
\def\ln{\,\mbox{ln}\,}

\def\al{\alpha}
\def\be{\beta}

\def\de{\delta}

\def\Ga{\Gamma}

\def\vep{\varepsilon}

\def\al{\alpha}
\def\be{\beta}

\def\de{\delta}

\def\be{\begin{equation}}
\def\ee{\end{equation}}

\def\bea{\begin{eqnarray}}
\def\eea{\end{eqnarray}}

\begin{document}

\begin{center}
{\large\sc  Sp(2) Renormalization} \vskip 6mm

Peter M. Lavrov$\footnote{E-mail address: lavrov@tspu.edu.ru}$
 \vskip 6mm

 \ \ {\small\sl Department of Mathematical Analysis, \\Tomsk State
Pedagogical University,\\ 634061, Kievskaya St. 60, Tomsk, Russia}
\vspace{0,4cm}

\end{center}
\vskip 10mm

\begin{quotation}
\noindent {\large\bf Abstract.}  The renormalization of general
gauge theories on flat and curved space-time backgrounds is
considered within the Sp(2)-covariant quantization method.
We  assume the existence of a gauge-invariant and diffeomorphism
invariant regularization. Using the Sp(2)-covariant formalism one
can show that the theory possesses gauge invariant and diffeomorphism
invariant renormalizability to all orders in the loop expansion
and the extended BRST symmetry after renormalization is preserved.
The advantage of the Sp(2)-method compared to the standard
Batalin-Vilkovisky approach is that, in reducible theories, the
structure of ghosts and ghosts for ghosts and auxiliary fields is
described in terms of irreducible representations of the Sp(2)
group. This makes the presentation of solutions  to the master
equations in more simple and systematic way because they are  Sp(2)-
scalars.
\vskip 4mm

 \vskip 2mm


\vskip 2mm

\noindent
{\sl MSC-AMS:} \
81T15, \
81T20.  \

\end{quotation}
\vskip 12mm

\section{Introduction}
It is well known that Green's functions in Quantum Field Theory (QFT)
contain  divergences \cite{BSh,Coll}. Renormalization should be considered
as one of important issue of QFT especially in gauge theories which form a
basis for formulating modern theories of fundamental interactions
(electromagnetic, weak, strong and gravitational). In the first papers
by \ 't Hooft and Veltman \cite{tHV1,tHV2} devoted to solving the problem
of renormalization in the Yang-Mills theories within Faddeev-Popov
quantization \cite{FP}, this achievement required a great effort.
In particular, it was necessary the invention of construction of
special gauges and also special technics to prove the gauge invariant
renormalizability. Later on, after deriving the Slavnov-Taylor identity
\cite{S}, discovering BRST symmetry \cite{brst} and presenting this
symmetry in the form of non-linear unique  equation (Zinn-Justin
equation) \cite{Z-J}, the proof of the gauge invariant renormalizability
of Yang-Mills theories  became more simple \cite{K-SZ,K}.

After discovering  supergravity theories 
\cite{FvNF,DZ,FvN} it was realized that direct application of the 
Faddeev-Popov answers leads in the case of these theories to an
incorrect result; namely, the violation of the physical $S$-matrix
unitarity. The reason lies in the structure of gauge transformations
for these theories.
In this case, the invariance transformations for the initial action
do not form a gauge group. The arising structure coefficients
may depend on the fields of the initial
theory, and the gauge algebra of these transformations may be opened by
terms proportional to the equations of motion. Moreover, attempts
of covariant quantization of  gauge theories with linearly-dependent
generators of gauge transformations result in the understanding of the
fact that it is impossible to use the Faddeev-Popov rules to construct
a suitable quantum theory
\cite{Town,S1,HKO}. Therefore, the quantization of gauge
theories requires taking into account many new aspects
(in comparison with
QED) such as open algebras, reducible generators and so on.
It was realized how to quantize them using different types of ghosts,
antighosts, ghosts for ghosts (Nielsen, Kallosh ghosts etc)
\cite{dWvH,FT,Nie1,Kal,Nie2,DeAGM,FrS,LT}.

A unique closed approach to the problem of covariant quantization
summarized all these attempts was proposed by Batalin and
Vilkovisky \cite{BV1,BV2}. The Batalin-Vilkovisky (BV)
formalism gives the rules for the quantization of  general gauge
theories. Now it is known that using new concept of renormalizability
(beyond the Dyson criterion \cite{Dy,MS}) proposed in \cite{VT} (see, also \cite{VLT}) this 
formalism enables one to prove the gauge-invariant renormalizability 
of general gauge theories when all fields under consideration are quantum ones.
Later this point of view was supported by Gomis and Weinberg \cite{GomisWein}
(see also \cite{HT,ABRKT,PS} for an extensive review and further references).

Renormalizations in curved space-time within the Dyson criterion are under 
intense investigations beginning with paper by Utiyama and DeWitt \cite{UDeW}
(see \cite{buch84,Toms,Panangaden,book,PoImpo} and references therein).
In the present work we continue our recent investigation of gauge
invariant renormalizability in curved space-time with the help of new concept of 
renormalizability \cite{VT}. In \cite{LSh} it was done in the framework of BV formalism 
\cite{BV1,BV2}.
We have extended these considerations to the case when the QFT is
defined in the presence of external conditions, in particular in
curved space-time and proved that in this case the gauge invariant
renormalizability is compatible with preserving general covariance.

Except the BV formalism, there is an alternative approach for
quantization of general gauge theories, which is based on the
principle of invariance under extended BRST symmetry including
BRST and anti-BRST transformations on an equal footing
\cite{BLT1,BLT2,BLT3} (compare with alternative approach  \cite{Hull}). 
We are going to consider the problem of
gauge invariant renormalizability of general gauge theories within the
framework of Sp(2)-method in the presence of a gravitational background field 
and to prove general covariance of renormalization.

The paper is organized as follows. In Section 2 an  exposition
of Sp(2) quantization approach in Lagrangian formalism for general
gauge theories is given.  In Section 3 within Sp(2) formalism the
general gauge theories in the presence of an external
gravitational field are considered. In Section 4 general
covariance of renormalization in the Sp(2) method is proved. In
Section 5 concluding remarks are given.

We use the condensed notations  as given by DeWitt \cite{DeWitt}.
Derivatives with respect to sources and antifields are taken
from the left, and those  with respect to fields, from the right.
Left derivatives with respect  to fields  are  labeled  by  the
subscript  {\it ``l''}.  The
Grassmann parity of any quantity $A$ is denoted by $\epsilon (A)$.

\section{Gauge theories in Sp(2)-covariant method}

In this section we present a very brief review of the
Sp(2)-covariant formalism \cite{BLT1,BLT2,BLT3}, which will be
used in the rest of the paper to prove the gauge invariant and
general covariant renormalizability of the quantum field theory on
curved background.

\subsection{Configuration space}

To construct the $Sp(2)$-quantization for general gauge theory one
needs in introduction of configuration space. To this end we
consider the initial classical action $S_0(A)$ of fields $A^i$.
This  action $S_0(A)$ is assumed to have at least one stationary
point $A_0=\{A^i_0\}$
\begin{eqnarray}
\label{EOMClassA}
 S_{0,i}(A)|_{A_0}=0,\quad S_{0,i}=\frac{\partial S_0}{\partial A^i}
\end{eqnarray}
and to be regular in the neighborhood of $A_0$. Equation
(\ref{EOMClassA}) defines a surface $\Sigma$ in space of functions
$A^i$. Invariance of the action $S_0(A)$ under the gauge
transformations $\delta A^i=R^i_{\alpha}(A) \xi^{\alpha}$ in the
neighborhood of the stationary point is assumed:
\begin{eqnarray}
\label{GIClassA}
 S_{0,i}(A) R^i_{\alpha}(A)=0,\quad
\alpha=1,2,...,m,\quad 0<m<n,\quad \varepsilon(\xi^{\alpha})
=\varepsilon_{\alpha}.
 \end{eqnarray}
Here $\xi^{\alpha}$ are arbitrary functions of space-time
coordinates , and $R^i_{\alpha}(A)$ $(\varepsilon(R^i_{\alpha})=
\varepsilon _i + \varepsilon_{\alpha})$ are generators of gauge
transformations. We have also used DeWitt's condensed notations
\cite{DeWitt}, when any index includes all particular ones (space -
time, index of internal group, Lorentz index and so on). Summation
over repeated indices implies integration over continuous ones and
usual summation over discrete ones.

Then it is necessary to introduce the total configuration space
$\Phi^A$, which coincides, in fact, with the total configuration
space in the BV formalism \cite{BV1,BV2}, but there is difference in
arrangement of the ghost and antighost fields:
\begin{eqnarray}
\label{ConfSpaceSp} \Phi^A = (A^i,\;B^{\alpha|a_1\cdot\cdot\cdot
a_s}, \;C^{\alpha |a_0\cdot\cdot\cdot a_s},\; s=0,...,L;\; a_i
=1,2),\quad \varepsilon (\Phi^A) = \varepsilon_A,
\end{eqnarray}
where $L$ denotes the stage of initial action reducibility.
 Auxiliary fields $B^{\alpha|a_1\cdot\cdot\cdot a_s}$ and
ghost fields $C^{\alpha |a_0\cdot\cdot\cdot a_s}$ are symmetric
Sp(2) tensors of corresponding ranks. The following values of the
Grassmann parity are ascribed to these fields: \bea
\nonumber
\vep(B^{\alpha|a_1\cdot\cdot\cdot a_s})&=&\vep_{\al_s}+s\;({\rm mod 2}),\\
\nonumber \vep(C^{\alpha |a_0\cdot\cdot\cdot a_s})&=&
\vep_{\al_s}+s+1\;({\rm mod 2}), \;\; s=0,...,L \eea together with
the following values of the ghost number: \bea
\nonumber gh(B^{\al_0})=0,\quad gh(B^{\alpha|a_1\cdot\cdot\cdot
a_s})=\sum^s_{s'=1}(3-2a_{s'}) ,\;\;\;  gh(C^{\alpha
|a_0\cdot\cdot\cdot a_s})= \sum^s_{s'=0}(3-2a_{s'}). \eea

To each field $\Phi^A$ of the total configuration space one
introduces three sets of antifields
$\Phi^*_{Aa},\;\varepsilon(\Phi^*_{Aa})=\varepsilon_A+1$ and
$\bar\Phi_A,\varepsilon(\bar\Phi_A)=\varepsilon_A$. We know the
meaning of antifields in the BV-approach. They are sources of BRST
transformations. In the extended BRST algebra, there are three
kinds of transformations; namely, BRST-transformations,
antiBRST-transformations and mixed transformations. The antifields
$\Phi^*_{Aa}$ form $Sp(2)$ doublets with respect to the index $a$
and can be treated as sources of BRST- and
antiBRST-transformations, while $\bar\Phi_A$ are sources of
combined transformation.

\subsection{Extended antibrackets}

On the space of fields $\Phi^A$ and antifields $\Phi^*_{Aa}$ one
defines odd symplectic structures $(\;,\;)^a$, called the 
extended antibrackets
\begin{eqnarray}
\label{AntiBSp} (F,G)^a\equiv\frac{\delta
F}{\delta\Phi^A}\;\frac{\delta G}{\delta\Phi^*_{Aa}}
-(F\leftrightarrow
G)\;(-1)^{(\varepsilon(F)+1)(\varepsilon(G)+1)}.
\end{eqnarray}
As usually the derivatives with respect to fields are understood
as acting from the right and those with respect to antifields, as
acting from the left.

The extended antibrackets (\ref{AntiBSp}) have the following
properties:
\begin{eqnarray}
\nonumber \label{PrAntiBSp}
&&\varepsilon((F,G)^a)=\varepsilon(F)+\varepsilon(G)+1,\\
\nonumber
&&(F,G)^a=-(G,F)^a(-1)^{(\varepsilon(F)+1)(\varepsilon(G)+1)},\\
\nonumber
&&(F,GH)^a=(F,G)^aH+(F,H)^aG(-1)^{\varepsilon(G)\varepsilon(H)},\\
&&
{((F,G)^{\{a},H)^{b\}}(-1)^{(\varepsilon(F)+1)(\varepsilon(H)+1)}
+{\rm cycl.perm.} (F,G,H)}\equiv 0,
\end{eqnarray}
where curly brackets denote symmetrization with respect to the
indices $a,b$ of the $Sp(2)$ group:
\begin{eqnarray}
A^{\{a}B^{b\}}\equiv A^aB^b + B^bA^a. \nonumber
\end{eqnarray}

The last relations in (\ref{PrAntiBSp}) are the graded Jacobi
identities for the extended antibrackets. In particular, for any
bosonic functional $S,\;\varepsilon(S) =0$, one can establish that
\begin{eqnarray}
\label{SSS} ((S,S)^{\{a},S)^{b\}} \equiv 0.
\end{eqnarray}

\subsection{Extended quantum master equations}

In addition the operators $V^a,\;\Delta^a$ are introduced
\begin{eqnarray}
\label{VaSp} V^a=\varepsilon ^{ab}\;\Phi^*_{Ab}\;
\frac{\delta}{\delta\bar\Phi_A},\;\;\;
\Delta^a=(-1)^{\varepsilon _A}\frac{\delta_{\it l}}{\delta\Phi^A}
\frac{\delta}{\delta\Phi^*_{Aa}},
\end{eqnarray}
where $\varepsilon ^{ab}$ is the antisymmetric tensor for raising
and lowering $Sp(2)$-indices
\begin{eqnarray}
\varepsilon ^{ab}=-\varepsilon ^{ba},\quad\varepsilon ^{12}=1\quad
\varepsilon_{ab}=-\varepsilon^{ab}. \nonumber
\end{eqnarray}

It can be readily established that the algebra of the operators
(\ref{VaSp}) has the form
\begin{eqnarray}
\label{AlgebraDeltaaSp} \Delta^{\{a}\Delta^{b\}}=0,\;\;\;
\Delta^{\{a}V^{b\}}+ V^{\{a}\Delta^{b\}}=0,\;\;\;
V^{\{a}V^{b\}}=0.
\end{eqnarray}
The action of the operators (\ref{VaSp}) on a product of
functionals $F$ and $G$ gives
\begin{eqnarray}
\label{ADeltaaSp} \Delta^a (F\cdot G) = (\Delta^a F)\cdot G +
F\cdot(\Delta^a G) (-1)^{\varepsilon (F)}  +
(F,\;G)^a(-1)^{\varepsilon (F)},
\end{eqnarray}
\begin{eqnarray}
\label{} \nonumber V^a(F,G)^b& =& (V^a F,\;G)^b -
(-1)^{\varepsilon (F)}
(F,V^a G)^b - \\
\nonumber &&- {\varepsilon}^{ab}\left(\frac{\delta
F}{\delta\phi^A} \frac{\delta G}{\delta\bar{\phi}_A} -\frac{\delta
G}{\delta\phi^A} \frac{\delta
F}{\delta\bar{\phi}_A}(-1)^{\varepsilon (F) (\varepsilon
(G)+1)}\right).
\end{eqnarray}
Therefore only the symmetrized form of $V^a$ acting on the
extended antibrackets observes the Leibniz rule
\begin{eqnarray}
\label{AVaSp} V^{\{a}(F,G)^{b\}} = (V^{\{a}F,G)^{b\}} -
(-1)^{\varepsilon (F)} (F,V^{\{a}G)^{b\}}.
\end{eqnarray}
For any bosonic functional $S$ we have
\begin{eqnarray}
\label{VSS} \frac{1}{2}V^{\{a}(S,S)^{b\}}=(V^{\{a}S,S)^{b\}}.
\end{eqnarray}
It is advantageous to introduce an operator $\bar\Delta^a$
\begin{eqnarray}
\nonumber \bar\Delta^a=\Delta^a+\frac{i}{\hbar}V^a
\end{eqnarray}
with the properties
\begin{eqnarray}
\label{AlgBarDeltaSp} \bar\Delta^{\{a}\bar\Delta^{b\}}=0.
\end{eqnarray}

For a boson functional $S=S(\Phi,\Phi^*,\bar\Phi)$, we introduce
extended quantum master equations
\begin{eqnarray}
\label{L1}
\frac{1}{2}(S,S)^a+V^aS=i\hbar\Delta^a S
\end{eqnarray}
with the boundary condition
\begin{eqnarray}
\label{L2} S\bigg|_{\Phi^* = \bar\Phi = \hbar = 0} = S_0(A),
\end{eqnarray}
where $S_0(A)$ is the initial classical action.

The generating equation for the bosonic functional $S$ is a set of
two equations. It should be verified that these equations are
compatible. The simplest way to establish this fact is to rewrite
the extended master equations in an equivalent form of linear
differential equations
\begin{eqnarray}
\label{L3} \bar\Delta^a \exp\left\{\frac{i}{\hbar}S\right\}=0.
\end{eqnarray}
Due to the properties of the operators $\bar\Delta^a$
(\ref{AlgBarDeltaSp}), we immediately establish the compatibility
of the equations.

\subsection{Gauge fixing}

The action $S$ is gauge-degenerate. To lift the degeneracy, we
should introduce a gauge. We denote the action modified by gauge
as $S_{ext}=S_{ext}(\Phi,\Phi^*,\bar{\Phi})$. The gauge should be
introduced so as, first, to lift the degeneracy in $\phi$ and,
second, to retain the extended master equation, which provides the
invariance properties of the theory for $S_{ext}$. To meet these
conditions, the gauge is introduced as
\begin{eqnarray}
\label{GF1}
 \exp\left\{\frac{i}{\hbar}S_{\it ext}\right\}=
\exp\left\{-i\hbar\hat{T}(F)\right\}\exp\left\{\frac{i}{\hbar}S\right\}
\end{eqnarray}
 where $F=F(\Phi)$ is a bosonic functional fixing a gauge in the theory.
 The explicit form of the operator $\hat{T}(F)$ is
\begin{eqnarray}
\label{GaugeOperSp} \hat{T}(F)=\frac{\delta
F}{\delta\Phi^A}\frac{\delta}{\delta\bar{\Phi}_A}
+\frac{i\hbar}{2}\varepsilon_{ab}\frac{\delta}{\delta\Phi^*_{Aa}}
\frac{\delta^2F}{\delta\Phi^A\delta\phi^B}\frac{\delta}{\delta\Phi^*_{Bb}}.
\end{eqnarray}
Due to the properties of the operators $\bar\Delta^a$, it is not
difficult to check the equality
\begin{eqnarray}
 \bar\Delta^a\exp\left\{-i\hbar\hat{T}(F)\right\}=
 \exp\left\{-i\hbar\hat{T}(F)\right\}\bar\Delta^a.
\end{eqnarray}
 Therefore, the action $S_{ext}$ satisfies the extended master
 equations
\begin{eqnarray}
 \bar\Delta^a\exp\left\{\frac{i}{\hbar}S_{\it ext}\right\}=0.
\end{eqnarray}

\subsection{Generating functional
 of Green's functions}

 We next define the generating functional $Z(J)$
 of Green's functions by the rule
\begin{eqnarray}
 Z(J)=\int d\Phi\exp\left\{\frac{i}{\hbar}
 [S_{\it eff}(\Phi)+J_A\Phi^A]\right\},
\end{eqnarray}
 where
\begin{eqnarray}
 S_{\it eff}=S_{ext}(\Phi,\Phi^*,\bar{\Phi})|_{\Phi^*=\bar{\Phi}=0}.
\end{eqnarray}
 It can be represented in the form
\begin{eqnarray}
\label{GFZSp}
 Z(J)&=&\int d\Phi\;d\Phi^{*}\;d\bar{\Phi}\;d\lambda\;d\pi^a\;\exp
 \bigg\{\frac{i}{\hbar}\bigg(S(\Phi,\Phi^{*},\bar{\Phi})+\Phi^{*}_{Aa}
 \pi^{Aa}+\nonumber\\&&
 +\bigg(\bar{\Phi}_A-\frac{\delta F}
 {\delta\Phi^A}\bigg)\lambda^A-\frac{1}{2}\varepsilon_{ab}\pi^{Aa}
 \frac{\delta^2F}{\delta\Phi^A\delta\Phi^B}\pi^{Bb}+
 J_A\Phi^A\bigg)\bigg\}\;,
\end{eqnarray}
 where we have introduced a set of auxiliary fields
 $\pi^{Aa}$, $\lambda^A$
\begin{eqnarray}
 \varepsilon(\pi^{Aa})=\varepsilon_A+1,\quad
 \varepsilon(\lambda^A)=\varepsilon_A.
\nonumber
\end{eqnarray}

\subsection{Extended BRST symmetry}

 An important property of the integrand for $J_A=0$ is its invariance
 under the following global transformations (which, for its part , is
 a consequence of the extended master equation for
 $S_{ext}$)
\begin{eqnarray}
 && \delta\Phi^A =\pi^{Aa}\mu_a,\quad
 \delta\Phi^*_{Aa}=\mu_a\frac{\delta S}{\delta\Phi^A},\quad
 \delta\bar{\Phi}_A=\varepsilon^{ab}\mu_a\Phi^*_{Ab},\nonumber\\
&& \delta\pi^{Aa}=-\varepsilon^{ab}\lambda^A\mu_b,\quad
 \delta\lambda^A=0,
\end{eqnarray}
 where $\mu_a$ is an Sp(2) doublet of constant anticommuting Grassmann
 parameters. These transformations realize the extended BRST
 transformations in the space of the variables $\Phi$, $\Phi^*$,
 $\bar{\Phi}$, $\pi$ and $\lambda$.

\subsection{Gauge independence of vacuum functional}

 The existence of these transformations enables one to establish the
 independence of the vacuum functional from the choice of gauge.
 Indeed, suppose $Z_F\equiv Z(0)$. We shall change the gauge $F\to F+
 \Delta F$. In the functional integral for $Z_{F+\Delta F}$ we make
 the above-mentioned change of variables with the parameters chosen as
\begin{eqnarray}
 \mu_a=\frac{i}{2\hbar}\varepsilon_{ab}\frac{\delta\Delta F}{\delta\Phi^A}
 \pi^{Ab}.
\end{eqnarray}
 Then we find
\begin{eqnarray}
 Z_F=Z_{F+\Delta F}
\end{eqnarray}
 and therefore the $S$-matrix is gauge-independent.

\subsection{Ward identities}

 Let us now derive the Ward identities, which follow from the fact
 that the boson functional $S(\phi,\phi^*,\bar{\phi})$ satisfies the
 extended master equations. To do this, we introduce the extended
 generating functional of Green's functions
\begin{eqnarray}
 {\cal Z}(J,\Phi^*,\bar{\Phi})=\int d\Phi\exp\left\{\frac{i}{\hbar}
 [S_{\it ext}(\Phi,\Phi^*_a,\bar{\Phi})+J_A\Phi^A]\right\}.
\end{eqnarray}
 From this definition it follows that
\begin{eqnarray}
 {\cal Z}(J,\Phi^*,\bar\Phi)|_{\Phi^*=\bar\Phi=0}=Z(J)
\end{eqnarray}
 where $Z(J)$ has been introduced above.

 We have,
\begin{eqnarray}
 \int d\Phi\;\exp\left\{\frac{i}{\hbar}J_A\Phi^A\right\}
 \bar{\Delta}^a
 \exp
 \left\{
 \frac{i}{\hbar}S_{ext}(\Phi,\Phi^*,\bar{\Phi})\right\}=0.
\nonumber
\end{eqnarray}
 Integrating by parts, under the assumption that the integrated
 expression vanishes, we can write this equality as
\begin{eqnarray}
\label{WIZSp} {\widehat\omega}^a{\cal Z}(J,\Phi^*,\bar{\Phi})=0,
\end{eqnarray}
where
\begin{eqnarray}
{\widehat\omega}^a=\bigg(J_A\frac{\delta}{\delta\Phi^*_{Aa}}-
\varepsilon^{ab}\Phi^*_{Ab}
 \frac{\delta}{\delta\bar{\Phi}_A}\bigg),\quad
{\widehat\omega}^{\{a}{\widehat\omega}^{b\}}=0.
\end{eqnarray}
Eqs. (\ref{WIZSp}) are the Ward identities for the generating
functional of Green's functions. For the generating functional
${\cal W}(J,\Phi^*,\bar{\Phi})$ of connected Green's functions we
have
\begin{eqnarray}
\label{WIWSp} {\widehat\omega}^a{\cal W}(J,\Phi^*,\bar{\Phi})=0,
\end{eqnarray}

 Finally, for the generating functional of vertex functions
\[
 \Gamma(\Phi,\Phi^*,\bar{\Phi})={\cal W}(J,\Phi^*,\bar{\Phi})-J_A\Phi^A,\;\;
 \Phi^A=\frac{\delta {\cal W}}{\delta J_A}
\]
 we obtain the Ward identities
\begin{eqnarray}
\label{WIGSp}
 \frac{1}{2}(\Gamma,\Gamma)^a+V^a\Gamma=0
\end{eqnarray}
 in the form of the classical part of the extended quantum
 master equations.

\subsection{Extended BRST invariant renormalizability}

Here  we  present the preservation of the extended BRST-symmetry
under renormalization within the usual assumptions on perturbation
theory as well as on a regularization repeating main arguments used
in \cite{VLT} to state the gauge invariant renormalizability in the
BV formalism.
It can be shown that if \footnote{The action of $\Delta^a
$-operators  on local functionals is proportional to $\delta(0)$.
Usually they say that a regularization (likes dimensional one) is
used when $\delta(0)=0$. Here a formal proof without using
this assumption is given.}
\begin{eqnarray}
&&\label{SGammaSp}
 \frac{1}{2}(S,S)^a+V^a S=i\hbar\Delta^a S,\\
&& \frac{1}{2}(\Gamma,\Gamma)^a+V^a\Gamma=0
\end{eqnarray}
 then the renormalized action $S_R$ and
the effective action $\Gamma_R$ satisfy the same equations
\begin{eqnarray}
\label{SGammaRSp}
 &&\frac{1}{2}(S_R,S_R)^a+V^a S_R=i\hbar\Delta^a S_R,\\
&& \frac{1}{2}(\Gamma_R,\Gamma_R)^a+V^a\Gamma_R=0.
\end{eqnarray}
(here and elsewhere we drop the index $ext$).

Let us  represent $S$ in the form
\begin{eqnarray}
\label{L4}
 S =\sum^{\infty}_{n=0}{\hbar}^n S_{(n)}=
S_{(0)} + \hbar S_{(1)} + \hbar^2 S_{(2)}+\cdot\cdot\cdot.
\nonumber
\end{eqnarray}
Then we have the following recurrent equations to define $S_{(n)}$
step by step beginning with $S_{(0)}$
\begin{eqnarray}
\label{S0} \frac{1}{2}(S_{(0)},\; S_{(0)})^{a}+V^a S_{(0)}= 0,
\end{eqnarray}
The $S_{(1)}$ and  $S_{(2)}$ satisfy the following linear
equations:
\begin{eqnarray}
\label{L5} (S_{(0)},\; S_{(1)})^a+V^a S_{(1)} = i\Delta^a S_{(0)}.
\nonumber
\end{eqnarray}
\begin{eqnarray}
\label{L6} (S_{(0)},\; S_{(2)})^a+V^a S_{(2)} = i\Delta^a
S_{(0)}-\frac{1}{2}(S_{(1)},\; S_{(1)})^a. \nonumber
\end{eqnarray}
In general
\begin{eqnarray}
\label{Sn} (S_{(0)},\; S_{(n)})^a+V^aS_{(n)}=i\Delta^a S_{(n-1)}-
\frac{1}{2}\sum_{k=1}^{n-1}(S_{(k)},S_{(n-k)})^a,\;\; n=1,2,3,...
\end{eqnarray}
In papers \cite{BLT1,BLT2} the existence theorem for the equations
(\ref{S0}) has been proved in the form of Taylor series in the
antifields $\Phi^*_{Aa},{\bar\Phi}_{A}$. For the gauge theories
discussed above the solution to (\ref{S0}) in the lower order in
antifields can be presented as
\begin{eqnarray}
\nonumber \label{S1t}
&&S_{(0)}=S_0(A)+A^*_{ia}R^i_{\alpha}C^{\alpha\mid a}+ {\bar
A}_iR^i_{\alpha}B^{\alpha}-\varepsilon^{ab}C^*_{\alpha a\mid
b}B^{\alpha}+\\
\nonumber &&+\sum_{s=1}^L\Big(C^*_{\alpha_sa\mid
a_0...a_{s-1}}C^{\alpha_s\mid aa_0...a_{s-1}}+{\bar
C}_{\alpha_s\mid a_1...a_s}B^{\alpha_s\mid
a_1...a_s}-\varepsilon^{ab}C^*_{\alpha_s a\mid
ba_1...a_s}B^{\alpha_s\mid a_1...a_s}-\\
&&\;\;\;\;\;\;\;\;\;\;\;\;-\frac{s}{s+1}B^*_{\alpha_sa_0\mid
a_1...a_{s-1}}B^{\alpha_s\mid
a_0a_1...a_{s-1}}-\varepsilon^{ab}C^*_{\alpha_sa\mid
ba_1...a_{s-1}}B^{\alpha_s\mid
a_1...a_{s-1}}\Big)+\cdot\cdot\cdot.
\end{eqnarray}
It is important to note that the functional $S_{(0)}$ (\ref{S1t})
is by construction  a local functional  if one operates with the
gauge algebra underlying a given gauge theory described in terms
of gauge generators being local functions.

Equations (\ref{Sn}) can be presented in the form
\begin{eqnarray}
 \label{Wa}
W^aS_{(n)}=F^a_n,
\end{eqnarray}
where
\begin{eqnarray}
 \label{Wa1}
W^a&=&\frac{\delta
S_{(0)}}{\delta\Phi^A}\frac{\delta}{\delta\Phi^*_{Aa}}
+(-1)^{\epsilon_A}\frac{\delta
S_{(0)}}{\delta\Phi^*_{Aa}}\frac{\delta_l}{\delta\Phi^A}+V^a,\\
\label{Fn} F^a_n&=&i\Delta^a S_{(n-1)}-
\frac{1}{2}\sum_{k=1}^{n-1}(S_{(k)},S_{(n-k)})^a.
\end{eqnarray}
The structure and properties of equations (\ref{Wa}) formally
coincide with ones used in \cite{BLT1,BLT2} to prove the existence
theorem. Indeed, operators  $W^a$ obey the relations
\begin{eqnarray}
 \label{Wa2}
W^{\{a}W^{b\}}=0
\end{eqnarray}
as consequences of equations (\ref{S0}) and the properties of $V^a$
(\ref{AlgebraDeltaaSp}). It follows from ({\ref{Wa}})  and
({\ref{Wa2}}) that $F^a_n$ should satisfy the equations
\begin{eqnarray}
 \label{WFn}
W^{\{a}F_n^{b\}}=0.
\end{eqnarray}
To prove these equations let us consider the identity ({\ref{SSS}})
and rewrite it in the form
\begin{eqnarray}
 \label{SSS1}
(\hbox{$\frac{1}{2}$}(S,S)^{\{ a}+V^{\{a}S-i\hbar \Delta^{\{a}S,S )^{b\}}-
(V^{\{ a}S-i\hbar \Delta^{\{a}S,S)^{b\}}=0,
\end{eqnarray}
or
\begin{eqnarray}
 \label{SSS2}
(S,\hbox{$\frac{1}{2}$}(S,S)^{\{ a}+V^{\{a}S-i\hbar \Delta^{\{a}S)^{b\}}+(V^{\{ a}S-i\hbar \Delta^{\{a}S,S)^{b\}}=0,
\end{eqnarray}
Using properties of operators $\Delta^a$ and $V^a$ (\ref{AlgebraDeltaaSp}), (\ref{VSS})
from (\ref{SSS2}) one derives
\begin{eqnarray}\nonumber
 \label{SSS3}
&&(S,\hbox{$\frac{1}{2}$}(S,S)^{\{ a}+V^{\{a}S-i\hbar \Delta^{\{a}S)^{b\}}+
V^{\{ a}[\hbox{$\frac{1}{2}$}(S,S)^{b\}}+V^{b\}}S-i\hbar \Delta^{b\}}S]-\\
&&\;\;\;\;\;\;\;\;\; -i\hbar \Delta^{\{a}[\hbox{$\frac{1}{2}$}(S,S)^{b\}}+V^{b\}}S-
i\hbar \Delta^{b\}}S]=0.
\end{eqnarray}
Note that
\begin{eqnarray}
 \label{SSS4}
\frac{1}{2}(S,S)^{a}+V^{a}S-i\hbar \Delta^{a}S=\hbar^n(W^aS_{(n)}-F^a_n)+O(\hbar^{n+1}),
\end{eqnarray}
in the lower order in $\hbar$ we have
\begin{eqnarray}
 \label{SSS5}
(S_{(0)},W^{\{a}S_{(n)}-F^{\{a}_n)^{b\}}+V^{\{a}[W^{b\}}S_{(n)}-F^{b\}}_n]=0,
\end{eqnarray}
or
\begin{eqnarray}
 \label{SSS6}
W^{\{a}[W^{b\}}S_{(n)}-F^{b\}}_n]=0
\end{eqnarray}
that proves (\ref{WFn}).
Repeating arguments given in
\cite{BLT1,BLT2} one can state the existence of solutions to the
Eqs. (\ref{Wa}) and therefore to (\ref{SGammaSp}). We suppose that
the action $S$ is a local functional.

Now let us  represent $\Gamma$ in the form
\begin{eqnarray}
\label{L7} \Gamma = S + \hbar ({\Gamma}^{(1)}_{div} +
{\Gamma}^{(1)}_{fin}) + \mbox{{\sl O}}({\hbar}^2)= S_{(0)}
+ \hbar ({\Gamma}^{(1)}_{div} + {\bar{\Gamma}}^{(1)}_{fin})
+ \mbox{{\sl O}}({\hbar}^2), \nonumber
\end{eqnarray}
where ${\bar{\Gamma}}^{(1)}_{fin}={\Gamma}^{(1)}_{fin} + S_{(1)}$.
Besides, ${\Gamma}^{(1)}_{div}$ and ${\Gamma}^{(1)}_{fin}$ denote
the divergent and finite parts of the one-loop approximation for
$\Gamma$.

The functional ${\Gamma}^{(1)}_{div}$ determines the counterterms
of the one-loop renormalized action $S_{1R}$ which  is the local
functional:
\begin{eqnarray}
S_{1R} = S - \hbar {\Gamma}^{(1)}_{div} \nonumber
\end{eqnarray}
and satisfies the equation
\begin{eqnarray}
\label{L8} (S_{(0)},\; {\Gamma}^{(1)}_{div})^a+V^a
{\Gamma}^{(1)}_{div}= 0. \nonumber
\end{eqnarray}
Let us consider
\begin{eqnarray}
\nonumber
&&\frac{1}{2}(S_{1R},\;S_{1R})^a+V^aS_{1R} - i\hbar\Delta^a S_{1R} = \\
\nonumber && =\frac{1}{2}(S,\;S)^a+V^aS - i\hbar\Delta^a S -\hbar
(S,\;{\Gamma}^{(1)}_{div})^a +
\frac{1}{2}{\hbar}^2({\Gamma}^{(1)}_{div},\;{\Gamma}^{(1)}_{div})^a
+
i{\hbar}^2\Delta^a {\Gamma}^{(1)}_{div} =  \\
\nonumber && ={\hbar}^2
\bigg(\frac{1}{2}({\Gamma}^{(1)}_{div},\;{\Gamma}^{(1)}_{div})^a +
i\Delta^a {\Gamma}^{(1)}_{div} -
(S_{(1)},\;{\Gamma}^{(1)}_{div})^a\bigg) + \mbox{{\sl O}}({\hbar}^3).
\end{eqnarray}
We find that $S_{1R}$ satisfies the master equation
\begin{eqnarray}
\label{L9} \frac{1}{2}(S_{1R},\;S_{1R})^a+V^aS_{1R} - i\hbar\Delta^a
S_{1R} = {\hbar}^2E_2^a + \mbox{{\sl O}}({\hbar}^3) \nonumber
\end{eqnarray}
up to certain terms $E_2^a$
\begin{eqnarray}
\label{L11} E_2^a =
\frac{1}{2}({\Gamma}^{(1)}_{div},\;{\Gamma}^{(1)}_{div})^a +
i\Delta^a {\Gamma}^{(1)}_{div} -
(S_{(1)},\;{\Gamma}^{(1)}_{div})^a \nonumber
\end{eqnarray}
of the second order in $\hbar$.

Let us construct the effective action ${\Gamma}_{1R}$ with the
help of the action $S_{1R}$. This functional is finite in the
one-loop approximation and satisfies the equation
\begin{eqnarray}
\label{L12} \frac{1}{2}({\Gamma}_{1R},\;{\Gamma}_{1R})^a+V^a
{\Gamma}_{1R}= {\hbar}^2E_2^a + \mbox{{\sl O}}({\hbar}^3). \nonumber
\end{eqnarray}
Represent ${\Gamma}_{1R}$ in the form
\begin{eqnarray}
\nonumber \label{L13} {\Gamma}_{1R}& = & S + \hbar
{\Gamma}^{(1)}_{fin} +
 + {\hbar}^2 ({\Gamma}^{(2)}_{1,div} +
{\Gamma}^{(2)}_{1,fin}) + \mbox{{\sl O}}({\hbar}^3)=\\
\nonumber &=&S_{(0)} + \hbar {\bar{\Gamma}}^{(1)}_{fin} +
 + {\hbar}^2 ({\Gamma}^{(2)}_{1,div} +
\bar{\Gamma}^{(2)}_{1,fin}) + \mbox{{\sl O}}({\hbar}^3),
\end{eqnarray}
where $\bar{\Gamma}^{(2)}_{1,fin}=\Gamma^{(2)}_{1,fin}+S_{(2)}$.
The divergent part ${\Gamma}^{(2)}_{1,div}$ of the two - loop
approximation for ${\Gamma}_{1R}$ determines the two - loop
renormalization for $S_{2R}$
\begin{eqnarray}
S_{2R} = S_{1R} - {\hbar}^2 {\Gamma}^{(2)}_{1,div} \nonumber
\end{eqnarray}
and satisfies the equations
\begin{eqnarray}
\label{14} (S_{(0)},\;
{\Gamma}^{(2)}_{1,div})^a+V^a{\Gamma}^{(2)}_{1,div} = E_2^a.
\nonumber
\end{eqnarray}
Let us now consider
\begin{eqnarray}
\nonumber
&&\frac{1}{2}(S_{2R},\;S_{2R})^a+V^aS_{2R} - i\hbar\Delta^a S_{2R} = \\
\nonumber &&=\frac{1}{2}(S_{1R},\;S_{1R})^a - i\hbar\Delta^a
S_{1R} - {\hbar}^2 (S_{1R},\;{\Gamma}^{(2)}_{1,div})^a +
i{\hbar}^3\Delta^a {\Gamma}^{(2)}_{1,div} =\\
\nonumber
&&={\hbar}^3\bigg(({\Gamma}^{(1)}_{div},\;{\Gamma}^{(2)}_{1,div})^a
+ i\Delta^a {\Gamma}^{(2)}_{1,div} -
(S_{(2)},{\Gamma}^{(1)}_{div})^a -
(S_{(1)},{\Gamma}^{(2)}_{1,div})^a\bigg) + \mbox{{\sl O}}({\hbar}^4)=\\
\nonumber &&={\hbar}^3E^a_3 + \mbox{{\sl O}}({\hbar}^4).
\end{eqnarray}
We find that $S_{2R}$ satisfies the master equations up to terms
$E^a_3$
\begin{eqnarray}
E^a_3=({\Gamma}^{(1)}_{div},\;{\Gamma}^{(2)}_{1,div})^a +
i\Delta^a {\Gamma}^{(2)}_{1,div} -
(S_{(2)},{\Gamma}^{(1)}_{div})^a -
(S_{(1)},{\Gamma}^{(2)}_{1,div})^a \nonumber
\end{eqnarray}
of the third order in $\hbar$. Then the corresponding effective
action $\Gamma_{2R}$ generated by $S_{2R}$ is finite in the two -
loop approximation
\begin{eqnarray}
\nonumber \label{16} {\Gamma}_{2R}& = &S + \hbar
{\Gamma}^{(1)}_{fin} + {\hbar}^2{\Gamma}^{(2)}_{1,fin}
 + {\hbar}^3 ({\Gamma}^{(3)}_{2,div} +
{\Gamma}^{(3)}_{2,fin}) + \mbox{{\sl O}}({\hbar}^4)=\\
\nonumber &=&S_{(0)} + \hbar {\bar{\Gamma}}^{(1)}_{fin} +
{\hbar}^2{\bar{\Gamma}}^{(2)}_{1,fin}
 + {\hbar}^3 ({\Gamma}^{(3)}_{2,div} +
{\bar{\Gamma}}^{(3)}_{2,fin}) + \mbox{{\sl O}}({\hbar}^4)
\end{eqnarray}
and satisfies the equations
\begin{eqnarray}
\label{L17}
\frac{1}{2}({\Gamma}_{2R},\;{\Gamma}_{2R})^a+V^a{\Gamma}_{2R}  =
{\hbar}^3E^a_3 + \mbox{{\sl O}}({\hbar}^4) \nonumber
\end{eqnarray}
up to certain terms $E_3$ of the third order in $\hbar$.

Applying the induction method we establish that the totally
renormalized action $S_R$
\begin{eqnarray}
S_R = S - \sum_{n=1}^{\infty}{\hbar}^n {\Gamma}^{(n)}_{n-1,div}
\end{eqnarray}
satisfies the quantum master equations exactly:
\begin{eqnarray}
\label{RMastEqBV} \frac{1}{2}(S_R,\;S_R)^a+V^a  = i\hbar\Delta^a
S_R,
\end{eqnarray}
while the renormalized effective action ${\Gamma}_R$ is finite in
each order of $\hbar$ powers:
\begin{eqnarray}
{\Gamma}_R = S + \sum_{n=1}^{\infty}{\hbar}^n
{\Gamma}^{(n)}_{n-1,fin}= S_{(0)} + \sum_{n=1}^{\infty}{\hbar}^n
{\bar{\Gamma}}^{(n)}_{n-1,fin},
\end{eqnarray}
and satisfies the identities
\begin{eqnarray}
\label{WIGammaRBV} \frac{1}{2}({\Gamma}_R,\;{\Gamma}_R)^a+V^a{\Gamma}_R = 0.
\end{eqnarray}
Here, we have denoted by ${\Gamma}^{(n)}_{n-1,div}$ and
${\Gamma}^{(n)}_{n-1,fin}$ the divergent and finite parts,
respectively, of the n - loop approximation for the effective
action which is finite in (n-1)th approximation and is constructed
from the action $S_{(n-1)R}$.

Therefore the renormalized action
$S_R$ and the effective action ${\Gamma}_R$ satisfy the quantum
master equations and the Ward identities, respectively. 
It is necessary to note that the Sp(2)-invariant renormalization was used in 
the paper \cite{TShS} to prove the conservation of new ghost number after renormalization. 
In what follows we will use the results of Sp(2)-invariant renormalization to prove the general covariance of renormalized generating functionals within Sp(2) formalism.

\section{General gauge theories in curved space within Sp(2) formalism}

Let us consider a theory of gauge fields $A^i$ in an
external gravitational field $g_{\mu\nu}$.
The classical theory is described by the action which
depends on both dynamical fields and external metric,
\beq
 S_0=S_0(A,g)\,.
\label{action}
\eeq
Here and below we use the condensed notation
$g\equiv g_{\mu\nu}$ for the metric, when it is an
argument of some functional or function. The action
(\ref{action}) is assumed to be gauge invariant,
\beq
 S_{0,i}R^i_a=0,\quad \delta A^i=R^i_a(A,g)\lambda^a
\,,\quad \lambda^a = \lambda^a(x)\quad (a=1,2,...,n)\,,
\label{gauge} \eeq as well as covariant, \beq \delta_g S_0 &=&
\frac{\delta S_0}{\delta A^i}\delta_g A^i + \frac{\delta S_0}{\delta
g_{\mu\nu}}\delta_g g_{\mu\nu}=0\,, \label{diff} \eeq where
$\lambda^a$ are independent parameters of the gauge transformation,
corresponding to the symmetry group of the theory. The
diffeomorphism transformation of the metric in Eq. (\ref{diff}) has
the form \beq \delta_g g_{\mu\nu} &=& -
g_{\mu\alpha}\partial_{\nu}\xi^{\alpha} -
g_{\nu\alpha}\partial_{\mu}\xi^{\alpha} -
\partial_{\alpha}g_{\mu\nu}\xi^{\alpha} \nonumber
\\
&=& - g_{\mu\alpha}\nabla_{\nu}\xi^{\alpha}
- g_{\nu\alpha}\nabla_{\mu}\xi^{\alpha}
\,=\, -\nabla_{\mu}\xi_{\nu}-\nabla_{\nu}\xi_{\mu}\,.
\label{xi}
\eeq
Here $\xi^\al$ are the parameters of the coordinates
transformation,
\beq
&& \xi^{\alpha} = \xi^{\alpha}(x)
\quad (\alpha=1,2,...,d)\,.
\eeq

The generating functional $Z(J,\Phi^*,\bar\Phi,g)$ of the Green
functions can be constructed in the form of the functional
integral \beq \label{Z} Z(J,\Phi^*,\bar\Phi,g)=\int d{\Phi}
\exp\Big\{\frac{i}{\hbar} \Big[S_{ext}(\Phi,\Phi^*,\bar\Phi,g)
+J_A{\Phi}^A\Big]\Big\}. \eeq Here $\Phi^A$ represents the full
set of fields of the complete configuration space of the theory
under consideration and $\Phi^*_{Aa},\bar\Phi_A$ are antifields.
Finally, $S_{ext}(\Phi,\Phi^*,\bar\Phi_A,g)$ is the quantum action
constructed with the help of the solution
$S=S(\Phi,\Phi^*,\bar\Phi_A,g) $ to the master equations \beq
\label{ME} \frac{1}{2}(S,S)^a+V^aS = i\hbar\Delta^aS\,, \qquad
S(\Phi,\Phi^*,\bar\Phi,g)|_{\Phi^*=\bar\Phi=\hbar=0} = S_0(A,g)
\eeq in the form given in Eqs. (\ref{GF1}), (\ref{GaugeOperSp}).
 Note that $S_{ext}$ satisfies the master
equations \beq
 \frac{1}{2}(S_{ext},S_{ext})^a+V^aS_{ext}=i\hbar\Delta^aS_{ext}.
\eeq

From gauge invariance of initial action (\ref{gauge}) in usual
manner one can derive the BRST symmetry and the Ward identities
for generating functionals $Z= Z(\Phi,\Phi^*,\bar\Phi,g), W=
W(\Phi,\Phi^*,\bar\Phi,g)$ and
$,\Gamma=\Gamma(\Phi,\Phi^*,\bar\Phi,g)$ in the form
(\ref{WIZSp}), (\ref{WIWSp}) and (\ref{WIGSp}) respectively.

In what follows we assume the general covariance of
$S=S(\Phi,\Phi^*,\bar\Phi,g)$, \beq \label{cov} \delta_g
S(\Phi,\Phi^*,\bar\Phi,g) = \frac{\delta S}{\delta \Phi^A}\delta_g
\Phi^A + \delta_g \Phi^*_{Aa}\frac{\delta S}{\delta \Phi^*_{Aa}} +
\frac{\delta S}{\delta g_{\mu\nu}}\delta_g g_{\mu\nu} = 0. \eeq

Let us choose the gauge fixing functional $F=F(\Phi,g)$ in a
covariant form \beq \delta_g F=0\,, \eeq then the quantum action
$S_{ext}=S_{ext}(\Phi,\Phi^*,\bar\Phi,g)$ obeys the general
covariance too \beq \label{covqa} \delta_g S_{ext}=0\,. \eeq

From the Eq. (\ref{covqa}) and the assumption that the term with
the sources $J_A$ in (\ref{Z}) is covariant \beq \label{covJ}
\delta_g (J_A\Phi^A) = (\delta_g J_A)\Phi^A+J_A(\delta_g\Phi^A) =
0\,, \eeq it follows the general covariance of
$Z=Z(J,\Phi^*,\bar\Phi,g)$. Indeed, \beq \nonumber\label{covZ}
&&\delta_g Z(J,\Phi^*,\bar\Phi,g) = \frac{i}{\hbar}\int d{ \Phi}
\Big[ \de_g \Phi^*_{Aa}\frac{\de
S_{ext}({\Phi},\Phi^*,\bar\Phi,g)} {\delta \Phi^*_{Aa}} +
\frac{\delta S_{ext}({\Phi},\Phi^*,\bar\Phi,g)} {\delta
g_{\mu\nu}}\delta_g g_{\mu\nu}+ \\
&&\quad +\de_g \bar\Phi_{A}\frac{\de
S_{ext}({\Phi},\Phi^*,\bar\Phi,g)} {\delta \bar\Phi_{A}}+(\delta_g
J_A){\Phi}^A\Big]\,
 \exp\Big\{\frac{i}{\hbar}\Big[ S_{ext}({\Phi},\Phi^*,\bar\Phi,g)
+ J_A{\Phi}^A\Big]\Big\}\,. \eeq

Making change of integration variables in the functional integral,
(\ref{covZ}), \beq
{\Phi}^A\;\;\;\rightarrow\;\;\;\Phi^A+\delta_g\Phi^A\,, \eeq we
arrive at the relation \beq \nonumber &&\delta_g
Z(J,\Phi^*,\bar\Phi,g) = \frac{i}{\hbar}\int d\Phi \Big[
\frac{\delta S_{ext}}{\delta \Phi^A}\delta_g \Phi^A + \delta_g
\Phi^*_A\frac{\delta S_{ext}}{\delta \Phi^*_A} + \delta_g
\bar\Phi_A\frac{\delta S_{ext}}{\delta \bar\Phi_A}+\frac{\delta
S_{ext}}{\delta g_{\mu\nu}}\delta_g g_{\mu\nu} + \\
\nonumber &&\quad +(\delta_g J_A)\Phi^A+J_A(\delta_g\Phi^A)\Big]
\exp\Big\{\frac{i}{\hbar}\Big[S_{ext}(\Phi,\Phi^*,\bar\Phi,g) +
J_A\Phi^A\Big]\Big\}
\\
&&\quad = \frac{i}{\hbar}\int d\Phi \Big[\delta_g S_{ext} +
\delta_g (J_A\Phi^A)\Big] \exp\Big\{\frac{i}{\hbar}
\Big[S_{ext}(\Phi,\Phi^*,\bar\Phi,g) +
J_A\Phi^A\Big]\Big\}\,=\,0\, \label{covZ1}. \eeq From
(\ref{covZ1}) it follows that the generating functional of
connected Green functions $W(J,\Phi^*,\bar\Phi,g))$ \beq
W(J,\Phi^*,\bar\Phi,g)=\frac{i}{\hbar}\ln Z(J,\Phi^*,\bar\Phi,g)
\eeq obeys the property of the general covariance as well \beq
\label{covW} \delta_g W(J,\Phi^*,\bar\Phi,g)=0\,. \eeq

Consider now the generating functional of vertex functions
$\Gamma=\Gamma(\Phi,\Phi^*,\bar\Phi,g)$ \beq
\Gamma(\Phi,\Phi^*,\bar\Phi,g)=
W(J,\Phi^*,\bar\Phi,g)-J_A\Phi^A\,, \eeq where \beq \label{Gamma}
\Phi^A=\frac{\delta W(J,\Phi^*,\bar\Phi,g)}{\delta J_A},\quad
J_A=- \frac{\delta \Gamma(\Phi,\Phi^*,\bar\Phi,g)}{\delta \Phi^A}.
\eeq From definition of $\Phi^A$ (\ref{Gamma}) and the general
covariance of $W(J,\Phi^*,\bar\Phi,g)$ we can conclude the general
covariance of $J_A\Phi^A $. Therefore, \beq \label{covGamma}
\delta_g\Gamma(\Phi,\Phi^*,\bar\Phi,g)= \delta_g
W(J,\Phi^*,\bar\Phi,g)=0, \eeq
the generating functional of vertex functions obyes the property of the 
general covariance too. So, in this Section it is proved that if an external 
gravitational background $g_{\mu\nu}$ does not destroy the gauge invariance of 
an initial action $S_0=S_0(A,g)$. then the generating functional of Green functions
can be constructed with the help of solution to the Sp(2)-master equations in an 
usual way. Moreover, if we assume the general covariance of the initial action then 
we prove the general covariance of non-renormalized generating functional of Green functions
as well as both the generating functional of connected Green functions and  
of vertex functions.

\section{Covariant renormalization in curved space-time}

Up to now we consider non-renormalized generating functionals of
Green functions. We are going to prove the general covariance for
renormalized generating functionals. For this end, let us first
consider the one-loop approximation for \
$\Gamma=\Gamma(\Phi,\Phi^*,\bar\Phi,g)$, 
\beq \label{Gamma1loop}
\Gamma =S + \hbar \big[
\Gamma^{(1)}_{div}+\Gamma^{(1)}_{fin}\big] + O(\hbar^2)\,, 
\eeq
where ${\bar\Gamma}^{(1)}_{div}$ and ${\bar\Gamma}^{(1)}_{fin}$
denote the divergent and finite parts of the one-loop
approximation for $\Gamma$. The divergent local term
$\Gamma^{(1)}_{div}$ gives the first counterpart in one-loop
renormalized action $S_{1R}$ \beq \label{S1loop}
S\rightarrow S_{1R} = S-\hbar\Gamma^{(1)}_{div}.
\eeq From (\ref{covqa}) and (\ref{covGamma}) it follows that in
one-loop approximation we have \beq \label{covGamma1loop} \delta_g
\big[\Gamma^{(1)}_{div}+\Gamma^{(1)}_{fin}\big] = 0 \eeq and
therefore $\Gamma^{(1)}_{div}$ and $\Gamma^{(1)}_{fin}$ obey the
general covariance independently \beq \label{covGamma1loops}
\delta_g\Gamma^{(1)}_{div}=0 \,,\qquad
\delta_g\Gamma^{(1)}_{fin}=0\,. \eeq

In its turn the one-loop renormalized action $S_{1R}$  is
covariant \beq \label{covS1loop} \delta_g S_{1R}=0\,. \eeq
Constructing the generating functional of one-loop renormalized
Green functions $Z_1(J,\Phi^*,\bar\Phi,g)$, with the action
$S_{1R}=S_{1R}(\Phi,\Phi^*,\bar\Phi,g)$, and repeating
arguments given above, we arrive at the relation \beq
\label{covF1loops} \delta_g Z_1=0 \,,\qquad \delta_g W_1 = 0
\,,\qquad \delta_g\Gamma_1=0\,. \eeq The generating functional of
vertex functions $\Gamma_1= \Gamma_1(\Phi,\Phi^*,\bar\Phi,g)$
which is finite in one-loop approximation \beq \label{Gamma2loop}
\Gamma_1 = S+\hbar\Gamma^{(1)}_{fin} + \hbar^2
\big[\Ga^{(2)}_{1,div} + \Ga^{(2)}_{1,fin}\big] + O(\hbar^3)\,,
\eeq contains the divergent part $\Gamma^{(2)}_{1,div}$ and
defines renormalization of the action $S$ in the two-loop
approximation \beq \label{S2loop} S\rightarrow S_{2R} =
S_{1R} - \hbar^2\Gamma^{(2)}_{1,div}\,. \eeq Starting from
(\ref{covGamma1loops}), (\ref{covS1loop}) and  (\ref{covF1loops})
we derive \beq \label{covGamma2loops} \delta_g\Gamma^{(2)}_{1,div}
\,=\, 0 \,,\qquad \delta_g\Gamma^{(2)}_{1,fin}=0\,, \eeq that
means general covariance of the divergent and finite parts of
$\Gamma_1$ in two-loop approximation. Therefore the two-loop
renormalized action $S_{2R}=S_{2R}(\Phi,\Phi^*,\bar\Phi,g)$
is covariant \beq \label{covS2loop} \delta_g S_{2R}=0. \eeq
\vskip 1mm

Applying the induction method we can repeat the procedure to an
arbitrary order of the loop expansion. In this way we prove that
the full renormalized action, $S_{R}=
S_{R}(\Phi,\Phi^*,\bar\Phi,g)$, 
\beq \label{SRloop}
 S_{R}\,=\,
S-\sum_{n=1}^{\infty}\hbar^n\Gamma^{(n)}_{n-1,div}\,, 
\eeq
which is local in each finite order in $\hbar$, obeys the general
covariance \beq \label{covSRloop} \delta_g S_{R}\,=\,0\,; \eeq
and the renormalized generating functional of vertex functions,
$\Gamma_R=\Gamma_R(\Phi,\Phi^*,\bar\Phi,g))$, \beq
\label{GammaRloop} \Ga_R\,=\,
S+\sum_{n=1}^{\infty}\hbar^n\Ga^{(n)}_{n-1,fin}\,, \eeq
which is finite in each finite order in $\hbar$, is covariant \beq
\label{covGammaRloop} \delta_g \Gamma_R\,=\,0\,. \eeq

Therefore, taking into account results of  Section 4 we can state
that in presence of an external gravitational field the gauge
invariant renormalizability  can be arrived with preserving
general covariance of functional $\Gamma$ (\ref{covGammaRloop}).

\section{Conclusions}

We have considered the general scheme of gauge-invariant and
covariant renormalization of the quantum gauge theories of matter
fields in flat and curved space-time. Using the Sp(2) formalism
we have proved that in the theory which admits gauge invariant and
diffeomorphism invariant regularization, these two symmetries hold
in the counterterms to all orders of the loops expansion together
with extended BRST symmetry. To arrive at these results we have used
the gauge invariant renormalizability of general gauge theories in
the Sp(2) formalism without assuming the use of regularization
for which acting by $\Delta^a $ on a local functional gives zero 
\cite{BLT3}. If one uses a regularization scheme where $\delta (0)=0$ then from 
the begining we have a solution $S_{(0)}$ to the classical master 
equations (\ref{S0}) (see \cite{BLT1}) and the Sp(2)-invariant 
renormalization  is given in the way described in Section 2 when 
$S_{(n)}=0, n=1,2,...$. 

\section*{Acknowledgments}
This paper was done during the visit of the author to Juiz de Fora,
supported by the FAPEMIG grant, project No.\ CEX-BPV-00045-10.
The author  is grateful to I.L. Shapiro and I.V. Tyutin for 
stimulating discussions,
to the Department of Physics of the Federal University of Juiz de
Fora for warm hospitality and to FAPEMIG for the grant.
The work was also supported by
the grant for LRSS, project No.\ 3558.2010.2, the  RFBR-Ukraine grant,
project No.\ 10-02-90446
and the RFBR grant, project No.\ 09-02-00078.



\end{document}